# Malthusian stagnation or Malthusian regeneration?

Ron W Nielsen aka Jan Nurzynski[1]

Environmental Futures Centre, Gold Coast Campus, Griffith University, Qld, 4222, Australia

October, 2013

Empirical evidence questions fundamental concepts of the human population dynamics. One of the key conclusions of this study is that positive checks activate the efficient Malthusian regeneration mechanism, suggesting that the Epoch of Malthusian Stagnation, the first stage of growth claimed by the Demographic Transition Theory, did not exist.

**Introduction[2]**

In the previous publication (Nielsen aka Nurzynski, 2013), we have discussed the postulate of the Epoch of Malthusian Stagnation, its origin and its many strongly questionable claims. We shall now investigate more closely the scientific merits of some of its fundamental concepts.

We have pointed out (Nielsen aka Nurzynski, 2013) that Malthus (1798) was aware of a possibility of two contrasting and probably exclusive mechanisms of growth activated by positive checks: growth-suppressing and growth-stimulating mechanisms – a mechanism that might be causing stagnation and a mechanism that might be causing regeneration and restoration of growth. Which of these two mechanisms is common in nature? Which of them is active now in the growth of human population and which of them might have been active in the past? Do positive checks suppress or stimulate growth? Did they create stagnation or regeneration? Did they suppress growth or did they activate an efficient replacement or regeneration mechanism as noticed by Malthus (1798) to repair quickly and efficiently any possible damage caused by positive checks?

---

[1] r.nielsen@griffith.edu.au; ronwnielsen@gmail.com; http://home.iprimus.com.au/nielsens/ronnielsen.html

[2] Discussion presented in this publication is a part of a broader study, which is going to be described in a book (under preparation): *Population growth and economic progress explained*.

**Fundamental concepts**

The postulate of the existence of the Epoch of Malthusian Stagnation, the first stage of growth proposed by the Demographic Transition Theory, is based on the following fundamental concepts:

1. For thousands of years, birth and death rates (fertility and mortality rates) were high and strongly fluctuating around their respective constant values, which were approximately the same for both of them.
2. Strong fluctuations in birth and death rates (fertility and mortality rates) were caused by positive checks (Malthus, 1798), i.e. by demographic crises: "subsistence crises, epidemic crises, combined crises (subsistence/epidemic), and finally crises from other causes, which are mainly exogenous (wars, natural or other catastrophes)" (Artzrouni & Komlos 1985, p. 24.).
3. Strong fluctuations were reflected in the strongly variable and unstable growth of human population manifested as a stagnant state of growth (Malthusian stagnation).
4. The Epoch of Malthusian Stagnation was created by the mechanism of Malthusian stagnation, i.e. by the growth-suppressing forces of positive checks.

The *first* fundamental concept can be considered as *unscientific* because it cannot be tested by data. *We have no data for the birth and death rates extending over thousands of years and we* *do not expect to have them.* We might have some circumstantial evidence but it is not enough to use it as the basis for a profound theory. Birth and death rates might have been high and strongly fluctuating but we cannot convincingly prove it, nor can we hope to prove it.

We cannot prove that fertility and mortality rates were high (Warf, 2010), that fertility was approaching its biological maximum (Omran, 2005), that "during the first stage [of the demographic transition], mortality vacillated at high levels "(Robine, 2001), that "the 'Age of Pestilence and Famine,' is characterized by high and fluctuating mortality rates" (McKeown, 2009, p. 20S.).

The *second* fundamental concept can be also regarded as *unscientific* for similar reasons. To prove the correlations between positive checks and the fluctuations in birth and death rates would be even harder than to prove that there were strong fluctuations. We have some data for demographic catastrophes but we cannot demonstrate that they were correlated with changes in birth and death rates.

We cannot prove that "variations in fertility and mortality [were] induced by epidemic, famine, and war" (Lee, 1997). We cannot prove that the vacillations of mortality at high level were caused *mainly* by infectious diseases and to a large extend by wars and famines (Robine, 2001), and that "epidemics, wars and famines [were repeatedly pushing] mortality levels to high peaks" (Omran, 2005, p. 733).

Even one fundamental but unscientific concept or assumption is not a good start for any theory but two make it even worse. For any theory to be acceptable, the fundamental assumptions should be as simple and as convincing as possible. The more we have to accept by faith the less reliable is the proposed explanation.

The *third* concept explains the *assumed* stagnant state of growth, by *assuming* that the fluctuations in the birth and death rates are reflected in the growth of human population, while ignoring the evidence (Shklovskii, 1962, 2002; von Foerster & Amiot, 1960; von Hoerner, 1975) that even though the growth was slow it was *not* stagnant.

The *fourth* fundamental concept is related to the first three concepts but it also stands well on its own. The *assumed* stagnant state of growth is explained as having been created by the Malthusian stagnation mechanism, the mechanism discussed in the previous publication (Nielsen aka Nurzynski, 2013).

We can test this concept by surveying demographic catastrophes and by studying their effects on the growth of human population. We can also study the population data (Maddison, 2010; Manning, 2008; US Census Bureau, 2013) to find out whether they display any clear signs of chaotic growth, which might be correlated with demographic crises and which might be described as stagnant. These two issues will be addressed in our next publications. The third way of testing this fourth concept is by studying the effects of adverse living conditions on the growth of human population, but first we shall investigate the effects of strong fluctuations in birth and death rates as well as the effects of the decreasing birth and death rates, the fundamental concepts of the first two stages of growth proposed by the Demographic Transition Theory.

**Fluctuations in birth and death rates**

As pointed out earlier (Nielsen aka Nurzynski, 2013) the size of human population is determined by the *difference* between the birth and death rates. To be more precise, and depending on their contribution, we should also include the immigration and emigration rates. However, in the first, and adequately accurate, approximation, the difference between birth and death rates determines the growth rate.

Constant growth rate produces *exponential* growth, while a zero growth rate produces a *constant size* of the population. Depending on their synchronisation, fluctuating birth and death rates may produce fluctuating growth rate. According to the concept of the Epoch of Malthusian Stagntion and of the Demographic Transition Theory, strong fluctuations in birth and death rates are reflected in the strongly variable growth of human population.

We might also consider the effects of the *averaged* growth rate, i.e. the growth rate obtained by averaging over the rapid fluctuations. Such process of averaging might produce a constant or slowly-oscillating growth rate. We now have two questions:

1. Is the growth of human population affected by strong fluctuations in the birth and death rates?
2. Is the growth of human population affected by the oscillating growth rate?

These questions should have been answered before proposing the existence of the Epoch of Malthusian Stagnation, the first stage of growth described by the Demographic Transition Theory, unless this theory is supposed to be taken just as an abstract paradigm dealing only with birth and death rates without being helpful in explaining the growth of human population.

It is interesting that, as pointed out earlier (Nielsen aka Nurzynski, 2013), the first question has been already answered *negatively* by computer simulations carried out first by Artzrouni and Komlos (1985) and later by Lagerlöf (2003), the simulations that were supposed to prove the existence of the Epoch of Malthusian Stagnation and of the transition to the second stage of growth. Contrary to the claimed evidence (Artzrouni & Komlos, 1985; Lagerlöf, 2003), close inspection of their results shows that the generated growth of population is neither stagnant nor oscillating but exponential and that the generated distributions do not fit the population data, results indicating that the assumed mechanism of the first two stages of growth proposed by the Demographic Transition Theory is incorrect.

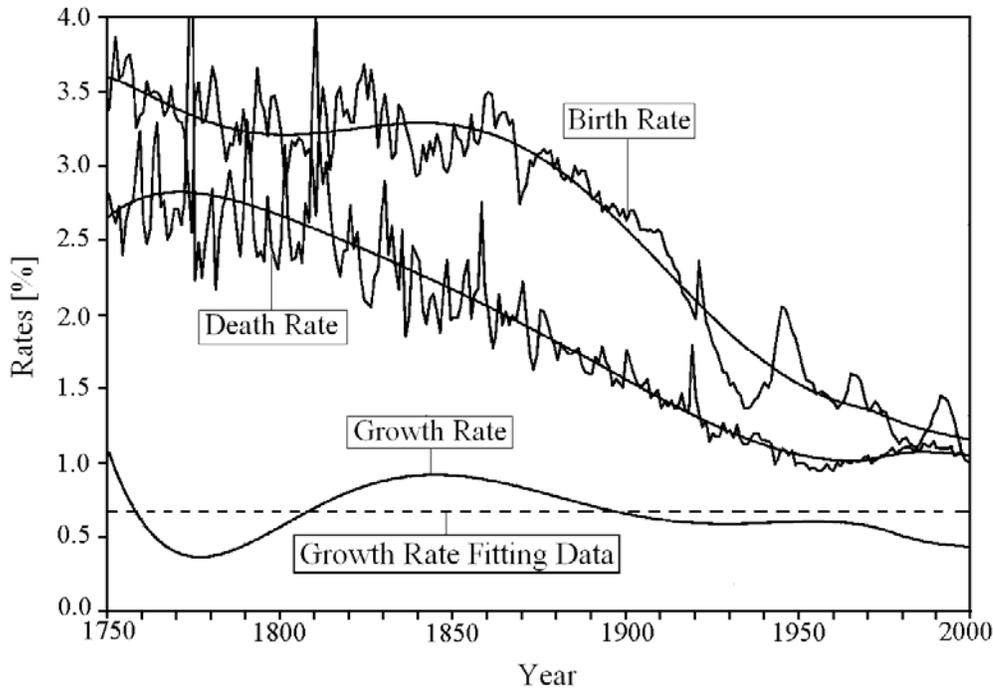

**Fig. 1.** Birth and death rates for Sweden between 1750 and 1998 (Statistics Sweden, 1999)

In order to demonstrate *empirically* that the growth of human population is *not* influenced by strong fluctuations in birth and death rates (fertility and mortality rates) or by moderate oscillations in the growth rate we can use the excellent data for the growth of the population in Sweden 1750-1998 (Statistics Sweden, 1999) because they contain the data for the birth and death rates and the corresponding data for the size of the population, making it unnecessary to use a tedious numerical integration to convert the data for the birth and death rates into the size of the population. These data are also particularly important because they are repeatedly used to demonstrate that the Demographic Transition Theory is confirmed by empirical evidence, while in fact they show that this theory does not work.

Fig. 1 appears to display all the hallmarks of the four stages of the Demographic Transition Theory:

(1) Stage 1: The high birth and death rates fluctuate strongly around approximately constant values.
(2) Stage 2: The widening gap between birth and death rates appears to indicate a transition to a new stage of growth. As required by the theory, birth and death rates gradually decrease but the decrease in birth rates is delayed. The fluctuations are also less violent and less frequent.
(3) Stage 3: The gap gradually becomes smaller indicating the third stage of growth.
(4) Stage 4: The gap starts to disappear suggesting the fourth stage of growth with a possibility of the birth rates becoming smaller than death rates and thus suggesting that Stage 4 might lead to Stage 5 (Haupt & Kane, 2005; Olshansky, Carnes, Rogers, & Smith, 1998; Schmid, 1984; van de Kaa, 2008)

Everything appears to look all right as long as we ignore the data describing the growth of population (Fig. 2). They show no signs of four stages of growth and, of course, no signs of transitions. The fundamental hallmarks of the Demographic Transition Theory suggested by the data for birth and death rates are not reflected in the growth of human population. These results suggest that we might be drawing incorrect conclusions when we study only birth and death rate, or even worse if we study just birth *or* death rates (fertility *or* mortality rates) while ignoring the aggregate data describing the size of the population.

In particular, by comparing Figs 1 and 2 we can see that the growth of human population between 1750 and 1850, i.e. when the birth and death rates were high and strongly fluctuating was neither stagnant nor chaotic. *Fluctuations in birth and death rates, even if strong and violent, do not have the slightest effect on the growth of human population.*

Fluctuations cannot possibly "testify to the continued existence of the 'Malthusian trap'" (Artzrouni & Komlos 1985, p. 24) because even strong fluctuations in death and birth rates do not produce any form of fluctuations in the size of human population and any form of stagnation. *Fluctuations do not influence the growth of human population.* The concept of the mechanism of stagnation and the existence of the Malthusian trap is not confirmed by the direct analysis of data, the data that according to the concept of the Epoch of Malthusian Stagnation should demonstrate the presence of such effects.

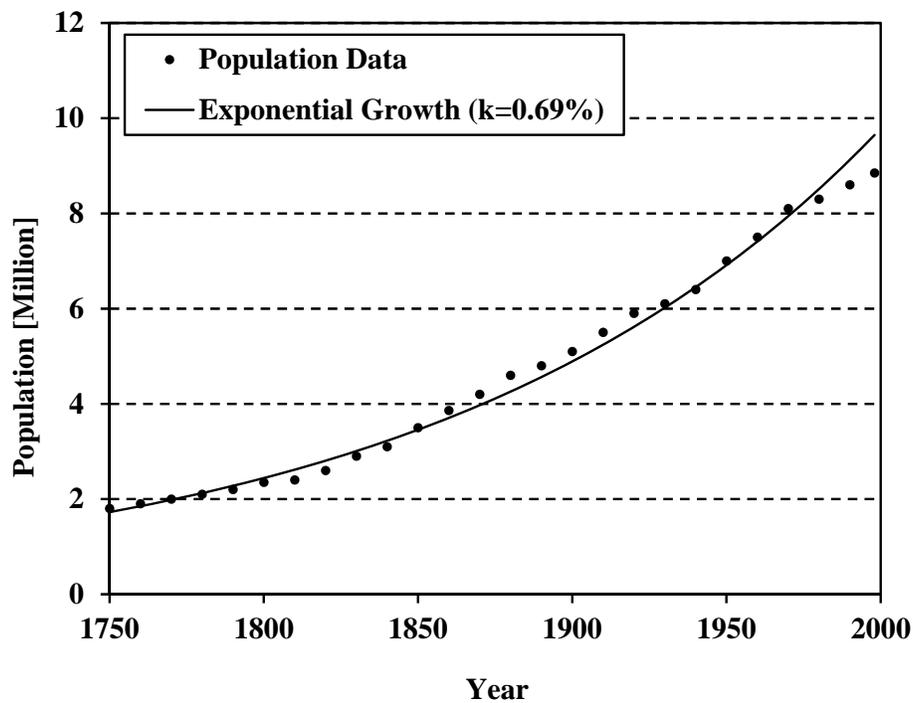

**Fig. 2.** Growth of population in Sweden, 1750-1998 (Statistics Sweden, 1999) fitted using exponential function with the growth rate of 0.69%.

We should also notice that during the time when the birth and death rates were high and strongly fluctuating as well as during the time when the average death and birth rates were decreasing and becoming less violent, the growth of population was following an undisturbed trajectory. The decreasing amplitude of fluctuations of death and/or birth rates does not indicate a transition to a new stage of growth. The decreasing average values of birth and/or death rates do not indicate a transition to a new stage of growth. The widening gap between the average values of death and birth rates does not necessarily indicate a transition to a new stage of growth.

The same applies to the narrowing of the gap. In order to determine whether there is a transition to a new stage of growth we would have to study carefully *how* the gap is widening or narrowing but more importantly we would have to identify an unusual change in the *size* of human population. If for instance, the size of the population, which was described satisfactorily using a certain mathematical distribution, *departs systematically* from this distribution, we may reasonably well conclude that there is a transition to a new stage of growth. However, it may be also only a temporary departure from the earlier trend or maybe we have used an incorrect mathematical description.

Studying only birth and death rates, while ignoring aggregate data describing the size of the population may be interesting but we might learn little or nothing from them about the growth of human population. Such a study alone might be even misleading. In order to understand the growth of human population we should study the growth of human population or at least to include it in our study.

Trying to explain the growth of human population by studying only birth and death rates is like trying to explain human behaviour by, for instance, studying only human metabolic reactions. Some of them might influence the way humans interact with the environment but if we want to understand human behaviour we should study the phenomena that are strongly reflected in human behaviour. The same applies to the study of the growth of human population.

The lower part of Fig. 1 shows the *difference* between the average values of birth and death rates, which determines the growth rate. We can see that the growth rate was not constant but that it was oscillating significantly around a constant value of 0.69%. We can certainly see clearly these oscillations. However, if we look at Fig. 2, we can see that these substantial variations in the growth rate are reflected surprisingly weakly in the size of the population. We can hardly see them. We can notice them only because we are comparing the data with the calculated exponential distribution corresponding to the constant growth rate of 0.69%.

Even significant variations in the growth rate (in the difference between average values of birth and death rates) may at best have only a negligible effect on the growth of human population. The growth of human population is fairly immune to any strong variations in the birth and death rates and to any reasonably strong variations in the growth rate. The growth of human population is determined primarily by the long-term changes in the growth rate.

If over a long time the growth rate is approximately constant, the growth will be exponential. If over a long time the growth rate can be described by a monotonically increasing distribution the growth of human population will be monotonically increasing but non-exponential, unaffected by any temporary fluctuations in the growth rate.

**Effects of positive checks**

We have pointed out earlier (Nielsen aka Nurzynski, 2013) that according to Malthus (1798) positive checks include not only large demographic catastrophes such as famines, pestilence and wars but also various forms of adverse living conditions. Impact of positive checks can be, therefore, studied by investigating how the parameters related to the living conditions are correlated with parameters describing the growth of population. This investigation is described in the forthcoming book mentioned in the footnote to the *Introduction* but we are presenting here one example from this study, the dependence of the annual growth rate on the level of deprivation (Fig. 3).

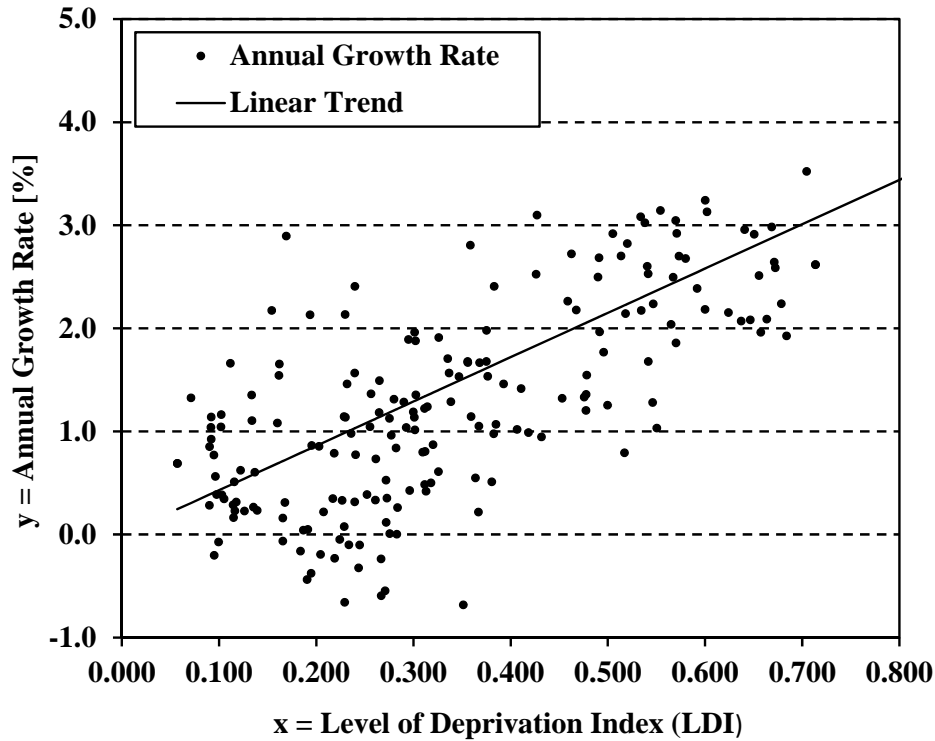

Fig. 3. The dependence of the annual growth rate on the level of deprivation based on the analysis of UNDP data (UNDP, 2011).

The level of deprivation is described by the Level of Deprivation Index (LDI), which we can defince simply by using the Human Development Index (HDI) (UNDP, 2010): LDI = 1 – HDI. High values of LDI, characterising third-world countries, are reflected in small ecological footprints, low per capita Gross Domestic Product, high levels of poverty, malnutrition, living with hunger, inadequate medical care, inadequate access to pure water, limited or no access to sanitation facilities, high death rates by polluted water, high level of maternal mortality, high rates of adult mortality and high rates of under-five mortality, all such circumstances as expected to reflect the assumed growth-suppressing effects of positive checks during the mythical Epoch of Malthusian Stagnation.

In contrast with the descriptions of the *assumed* growth-suppressing impacts of positive checks, empirical evidence shows the opposite effect: *strong positive checks stimulate growth*. The analysis of the UDNP data (discussed in the forthcoming book) shows that birth and death rates increase with the level of deprivation, in the apparent agreement with the concept of the Epoch of Malthusian Stagnation, but it also shows that the higher is the level of deprivation the higher is also *the annual growth rate* (see Fig. 3).

The difference between the *assumed* effects of strong positive checks and the empirical evidence is summarised in Table 1. Empirical evidence is in disagreement with the concept of the Epoch of Malthusian Stagnation and with the Demographic Transition Theory.

**Table 1.** Impacts of strong positive checks assumed by the Demographic Transition Theory are compared with empirical evidence

| **Assumed impact** | **Empirical evidence** |
|---|---|
| Birth and death rates *high* | Birth and death rates *high* |
| Growth rate *small* | Growth rate *large* |
| *Stagnant* state of growth | *Vigorous* state of growth |

Fig. 3 shows that the annual growth rate is directly proportional to the level of deprivation. The higher is the level of deprivation the larger is the growth rate. The fitted straight line to the data is given by following simple equation:

$$y = ax \qquad (1)$$

where $y$ is the annual growth rate in per cent, $x$ is the level of deprivation expressed as LDI and $a = 4.3\%$ for this particular set of data (UNDP, 2011).

The growth rate decreases to zero for the lowest possible level of deprivation but increases to 4.3% for the highest level. If positive checks are weak the growth rate is relatively small but it increases when the intensity of positive checks increases.

Empirical evidence strongly suggests that: (1) positive checks activate the efficient Malthusian replacement mechanism; (2) positive checks do not suppress the growth of human population but stimulate it; (3) the fundamental mechanism, the mechanism of Malthusian stagnation, did not work; (4) positive checks could not have created the Epoch of Malthusian Stagnation.

If the demographic crises (positive checks) in the distant past were strong, as assumed in the concept of the Epoch of Malthusian Stagnation and in the Demographic Transition Theory, they would have stimulated a fast growth of the population, but the population data (Maddison, 2010; Manning, 2008; US Census Bureau, 2013) show that the growth of population, global and regional, was slow and consequently that the growth rate was small. Based on these two arguments we now have the following conclusion (prediction): Demographic crises (positive checks) in the distant past were too weak to have any significant impact on the growth of human population.

We shall examine this prediction in the next publication. This prediction, if confirmed, contradicts the fundamental claim expressed in the concept of the Epoch of Malthusian Stagnation that high birth and death rates and the small growth rate (the gap between birth and death rates) is caused by demographic crises (positive checks).

**Conclusions**

The discussion presented in this publication leads to the following essential conclusions:

1. The concept of the existence of high and strongly fluctuating birth and death rates (fertility and mortality rates) in the distant past and over thousands of years may be regarded as *unscientific* because it cannot be ever expected to be verified by relevant data.

2. For the same reasons, the concept that strong fluctuations in birth and death rates (fertility and mortality rates) were correlated with (caused by) incidents of demographic crises may be also regarded as *unscientific*. Data of exceptionally high quality would be required to prove the existence of such close correlations but while we have some limited data about demographic catastrophes we have no data about birth and death rates extending over a long time in the distant past.

3. These two fundamental concepts of the existence of the Epoch of Malthusian Stagnation have to be either accepted by faith or rejected because they cannot be proven which means that the explanations offered by the concept of the Epoch of Malthusian Stagnation and by the Demographic Transition Theory have to be also accepted by faith or rejected.

4. Even strong fluctuations in birth and death rates (fertility and mortality rates) have *no effect on the growth of human population*. The hypothesis of the Epoch of Malthusian Stagnation starts with an unscientific concept of the existence of fluctuations and with another unscientific concept that the fluctuations were caused by demographic crises and continues with the empirically unsupported concept that the fluctuations are reflected in the growth of human population.

5. Even sizable oscillations in the growth rate have only small or negligible effect on the growth of human population.

6. The growth of human population is determined primarily by the long-term changes in the growth rate, i.e. by the long-term variations of the difference between birth and death rates.
7. The decreasing amplitude of the fluctuations in death and/or birth rates does not indicate a transition to a new stage of growth.
8. The decreasing average values of birth and/or death rates do not indicate a transition to a new stage of growth.
9. The widening gap between the average values of death and birth rates do not necessarily indicate a transition to a new stage of growth.
10. The annual growth rate increases with the increasing level of deprivation, i.e. with the intensity of positive checks. If positive checks are weak the growth rate is relatively small but it increases when the intensity of positive checks increases.
11. Empirical evidence strongly suggests that positive checks do not suppress the growth of human population but stimulate it.
12. Empirical evidence strongly suggests that positive checks could not have created the Epoch of Malthusian Stagnation.
13. Empirical evidence strongly suggests that the fundamental mechanism, the mechanism of Malthusian stagnation, did not work.
14. Empirical evidence strongly suggests that positive checks activate the efficient Malthusian replacement/regeneration mechanism.
15. Demographic crises (positive checks) in the distant past were too weak to have any significant impact on the growth of human population. This conclusion (prediction) has to be checked by studying the records of demographic catastrophes.

The study of fluctuations in birth and death rates contradicts the concept of the Epoch of Malthusian Stagnation. Linking them with incidents of demographic crises, such as famines, pestilence and wars, and claiming that such crises were suppressing the growth of human population and that fluctuations are the testimony to the existence of Malthusian trap creating the stagnant state of growth are not supported by the direct examination of empirical evidence. The postulate that demographic crises were suppressing the growth of human population is contradicted by the examination of the dependence of the annual growth rate on the level of deprivation, which shows that the annual growth rate increases with the level of deprivation, i.e. that the effect is the opposite to that claimed by the concept of the Epoch of Malthusian Stagnation but in perfect agreement with the concept of the efficient regeneration mechanism suggested first by Malthus (1798). The stronger are the forces, which are supposed to be suppressing the growth, the stronger is also the force of procreation. Paradoxically, perhaps, and in contradiction with the concept of the Epoch of Malthusian Stagnation, empirical evidence indicates that adverse living condition stimulate growth and that the Malthusian stagnation mechanism does not work.

We have discussed here two ways of showing that the concept of the Epoch of Malthusian Stagnation and that the associated concepts of the Demographic Transition theory are unsupported by empirical evidence but we can reinforce this conclusion by further examination of other types of data, by a closer investigation of the effects of demographic catastrophes and by the investigation of the population growth trajectories.